# Ultrashort pulse generation from a graphene-coupled passively mode-locked terahertz laser

Elisa Riccardi,[1][*] Valentino Pistore,[1][*] Seonggil Kang,[2] Lukas Seitner,[3] Anna De Vetter,[2] Christian Jirauschek,[3] Juliette Mangeney,[2] Lianhe Li,[4] A. Giles Davies,[4] Edmund H. Linfield,[4] Andrea C. Ferrari,[5] Sukhdeep S. Dhillon,[2] and Miriam S. Vitiello[1]

[1] *NEST, CNR - Istituto Nanoscienze and Scuola Normale Superiore, Piazza San Silvestro 12, 56127, Pisa, Italy*

[2]*Laboratoire de Physique de l'Ecole Normale Supérieure, ENS, Université PSL, CNRS, Sorbonne Université, Université de Paris, Paris, France*

[3]*Department of Electrical and Computer Engineering, Technical University of Munich, Hans-Piloty-Str. 1, 85748 Garching, Germany*

[4] *School of Electronic and Electrical Engineering, University of Leeds, Leeds LS2 9JT, UK*

[5]*Cambridge Graphene Centre, University of Cambridge, Cambridge CB3 0FA, UK*

[*]*These authors contributed equally to this work*

**The generation of stable trains of ultra-short (fs–ps), terahertz (THz)-frequency radiation pulses, with large instantaneous intensities, is an underpinning requirement for the investigation of light-matter interactions, for metrology and for ultra-high-speed communications. In solid-state electrically-pumped lasers, the primary route for generating short pulses is through passive mode-locking. However, this has not yet been achieved in the THz range, defining one of the longest standing goals over the last two decades. In fact, the realization of passive mode-locking has long been assumed to be inherently hindered by the fast recovery times associated with the intersubband gain of THz lasers. Here, we demonstrate a self-starting miniaturized ultra-short pulse THz laser, exploiting an original device architecture that includes the surface patterning of multilayer-graphene saturable absorbers distributed along the entire cavity of a double-metal semiconductor 2.30–3.55 THz wire laser. Self-starting pulsed emission with 4.0-ps-long pulses in a compact, all-electronic, all-passive and inexpensive configuration is demonstrated.**

In laser physics, mode-locking is a fundamental set of techniques that allows the generation of ultra-short light pulse trains with large instantaneous intensities from a laser source[1–3]. Owing to their remarkable stability[4], high peak-power[5,6], ultra-short duration[7,8] and coherent emission over a broad bandwidth[8], mode-locked laser sources benefit an exceptional variety of applications, including micro-[9] and nano-fabrication[10], optical data storage[11], microscopy[12], telecommunications[13], surgery[14] and research in ultrafast phenomena[15], among many others. Most notably, frequency combs (FC)[16] can be generated from stabilized mode-locked lasers[17], and hence are key for quantum metrology, sensing, communication, and spectroscopy[17].

Passive mode-locking techniques, relying on semiconductor saturable absorber mirrors (SESAM)[18] or nonlinear phase shifts[19–21], are commonly used to generate ultra-short pulses from the ultraviolet (from $\lambda \sim 200$nm[22]) to the mid-infrared (up to $\lambda \sim 3.5\mu$m[23]). At longer wavelengths such as terahertz (THz) frequencies ($\lambda \sim 30\mu$m – $300\mu$m), the lack of suitable gain media has resulted in the adoption of different strategies to generate pulsed laser radiation, either through optical rectification processes using femtosecond optical laser[24], although with limited (~$\mu$W) output power and efficiency ($\eta < 6\%$)[25,26], or requiring complex, non-tabletop and expensive setups, such as free electron lasers[27], hindering their exploitation beyond basic research.

Quantum cascade lasers (QCLs) are the only miniaturized direct sources of laser radiation in the THz frequency range, combining chip-scale size, high (> 1W) power emission, high spectral purity and broad bandwidth[28]. Although room temperature operation has still to be achieved, Fabry-Perot THz QCLs can operate in a compact Peltier cooler configuration up to 250 K[29] and can generate trains of short pulses by active mode-locking[30–32], i.e. by modulating the gain and losses in the active medium with a bias current modulation. Active techniques are favored by the QCL ultrafast carrier dynamics inherent to the intersubband transitions in the gain medium[33]. The gain in a QCL recovers from its saturated value much faster (~ 2 ps[34]) than the cavity round-trip time (~72 ps for a standard 3-mm-long cavity) and the photon lifetime[35]. This is the result of very strong electron-electron and electron-phonon interactions in the QCL polar semiconductor gain media, which induce an ultrafast non-radiative intersubband relaxation[36]. This, in turn, prevents the successful use of conventional passive mode-locking techniques based on lumped absorbers[37], since local perturbations of the photon population are quickly reabsorbed by the fast gain that quickly restores a quasi-continuous wave (CW) emission profile[38,39]. Recent approaches for short pulse emission in THz QCLs are therefore all based on active schemes. These include dispersion compensation of the active region, leading to the generation of a train of 4-ps-long pulses[30], or generation of shorter isolated pulses (2.5 ps)[32], albeit with a duration limited by the slow electrical modulation employed in the active modelocking[40]. The latter, which makes inefficient use of the gain, also requires dedicated external electronics and connections, impacting its widespread adoption.

Although self-starting short pulse emission from THz QCLs is highly desirable for spectroscopic[41], imaging[42], metrological[43] and Information and Communication Technologies (ICT) applications[44], and for providing a compact alternative to bulky THz time-domain systems[45], passive mode-locking in THz QCLs has not been reported to date, to the best of our knowledge. Theoretical predictions of possible design strategies (i.e. interleaved gain and absorbing periods with appropriate dipole moments)[46] have been followed by the encouraging experimental observation of Rabi-flopping in mid-IR QCLs[47]. However, at THz frequencies, engineering intra-cavity semiconductor multilayers with stringent requirements of the gain and absorption faces fundamental obstacles due to the extremely small photon energies involved.

Here we demonstrate passive mode-locking in a semiconductor heterostructure laser operating at THz frequencies. By employing a heterogeneous gain medium[48,49], and a cavity architecture that integrates a distributed graphene saturable absorber (DGSA) on the top-surface of the double-metal QCL cavity, we achieve self-starting pulsed emission with 4.0-ps-long pulses in a compact, all-electronic, all-passive and inexpensive configuration. We take advantage of the high transparency modulation (~80%)[50] and ultrafast recovery time (2–3 ps)[51–53] (i.e. faster than the gain recovery time in THz QCLs) of graphene saturable absorption, while the intracavity THz radiation experiences the saturable losses, thus favoring pulsed emission over the naturally occurring CW emission. Detailed electromagnetic simulations of the QCL structure and light-matter interactions in the active medium corroborate our findings, confirming the robustness of our approach. The proposed scheme can be applied to any semiconductor heterostructure laser, including mid-IR QCLs - with the first claims of passive mode-locking of QCLs operating in the mid-infrared[54] re-interpreted as coherent dynamic instabilities resulting from the extremely fast gain recovery time of this family of "class A" lasers[34,35,55].

**Results**

The QCL gain medium comprises 9 GaAs quantum wells, forming a cascade of alternating photon and LO-phonon-assisted transitions between two quasi-minibands[48,56]. It consists of a 17-μm-thick GaAs/AlGaAs heterostructure with 3 active regions having gain bandwidths centered at 2.5, 3 and 3.5 THz and comparable threshold current densities. The double-metal Fabry-Perot laser cavity includes top Ni-based lossy side absorbers, to suppress high-order lateral modes (Methods)[32]. A reference Fabry-Perot QCL 2.9-mm-long and 85-μm-wide was fabricated with a set of 6.5-μm-wide side absorbers (see Ref.[49] for transport and optical data), while a 2.3-mm-long and 49-µm-wide QCL with 0.5-µm-wide side absorbers was prepared for multilayer graphene (MLG) integration. Two 4-µm-wide stripes, aligned with the waveguide longitudinal axis and centered 12µm from each edge, were lithographically defined onto the top Au layer, exposing the active region below. The DGSA was realized by transferring a seven-layer CVD-grown MLG film on top of the etched stripes (Fig. 1A,B), (Methods). MLG on the Au contact outside the stripes is removed to allow application of the bonding wires (Fig. 1A), whereas the MLG ensures a homogeneous covering of the active region (Fig. 1B).

The DGSA-QCL was simulated via COMSOL Multiphysics. The MLG is included as a transition boundary condition with refractive index $ñ_g$ determined by THz time-domain spectroscopy (TDS) experiments (Tera K5 by MenloSystems) performed on the nominally same MLG film transferred onto an undoped GaAs substrate, resulting in $ñ_g = 16.9 + i\, 49.2$ at 3 THz (Supplementary Material). The highly doped n$^+$ layer (n = $2 \times 10^{18}$) between the active region and the top Au contact is also included in the model. Figure 1C is the electric field intensity distribution in a cross section of the QCL cavity at 3 THz. MLG affects the electric field distribution across the interface at the active region/Au top contact/air boundary at y = 17 µm, inducing the concentration of the electric field on the MLG. Refs[50,57] showed that a strong THz electric field can saturate the graphene absorption, $α_g$, which results in a decrease of the imaginary part of its complex refractive index $k_g$ due to its proportionality to the absorption coefficient $α_g = 4πk_g/λ_0$[58], where $λ_0$ is the radiation wavelength.

We then simulated the DGSA-induced waveguide losses spectra for the TM$_{00}$ mode at 3 THz for different values of the imaginary part of the MLG refractive index (Fig. 1D). For each frequency, the losses have a maximum, marked by the black dashed line in the map. Below this curve, a reduction of Im($ñ_g$) corresponds to lower losses, i.e. the MLG stripes behave as a DGSA. Im($ñ_g$), as extracted from THz-TDS, is displayed as the purple dashed line in Fig. 1D. This guarantees DGSA operation at least for all frequencies > 2.5 THz. The TM$_{00}$ and TM$_{01}$ waveguide losses decrease for stronger saturation, thus leading to a net round-trip gain for pulsed over continuous wave emission. The opposite would be true if Im($ñ_g$) were above the loss maximum. Figure 1E shows the TM$_{00}$ and TM$_{01}$ mode losses at 3 THz as a function of Im($ñ_g$). The starting Im($ñ_g$) corresponds to the orange dashed line, so the TM$_{01}$/TM$_{00}$ ratio is 55%–60% regardless of intracavity field intensity. The saturable and non-saturable absorption coefficients can also both be extracted for the TM$_{00}$ mode, each having an approximate value of 10.5 cm$^{-1}$, by comparing the losses at Im($ñ_g$)=49.2 and Im($ñ_g$)=0.

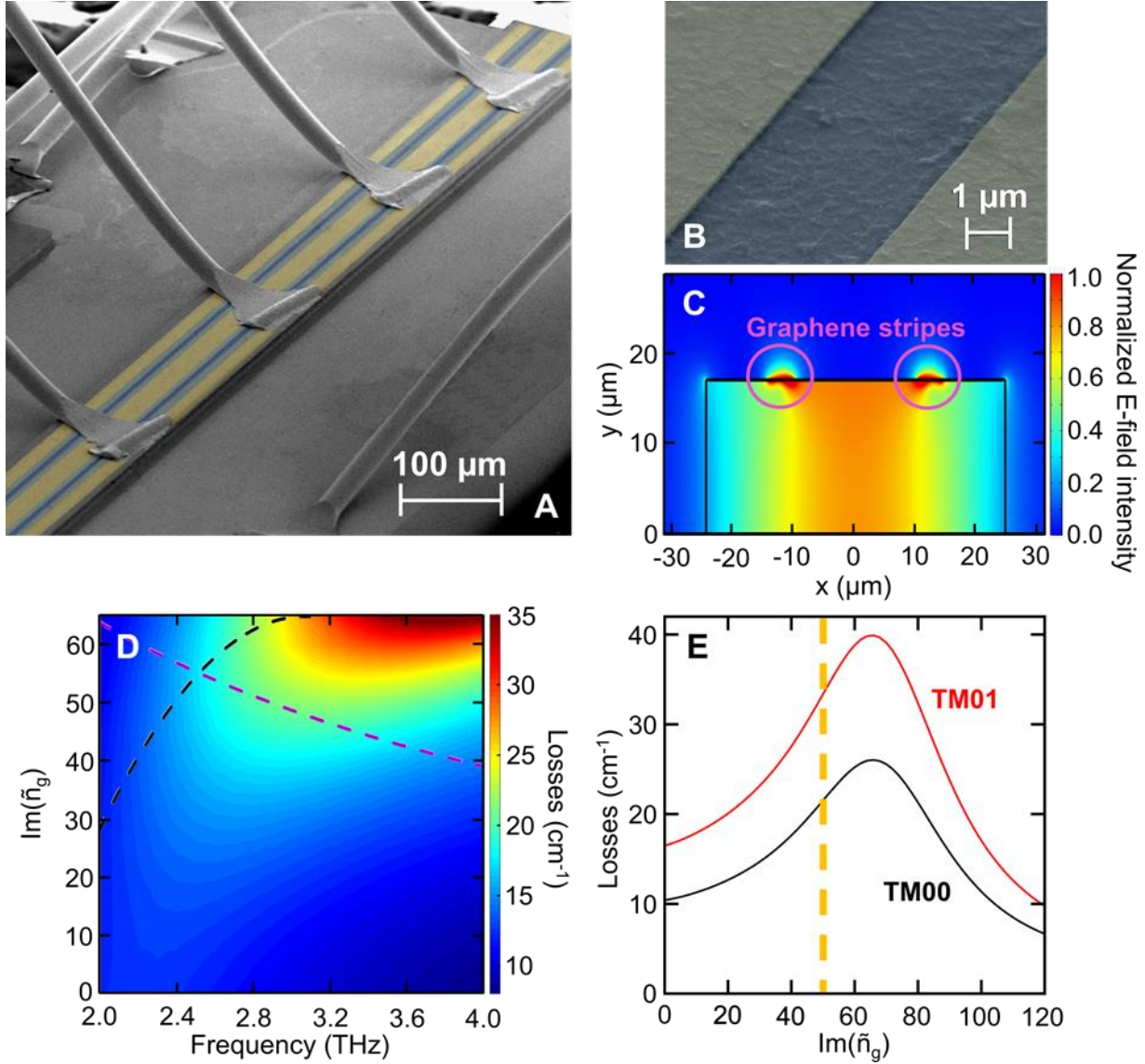

**Fig. 1. Device schematic and optical simulations**.
**A**, False color scanning electron microscope (SEM) image of a double metal QCL with an integrated DGSA. **B**, Magnified image over a portion of the QCL cavity, showing the Au top contact with the 4 μm etched stripes covered by the MLG (in blue). **C**, Simulated electric field intensity distribution in the cross-section of the DGSA-QCL at 3 THz. The violet circles highlight the MLG stripes. The MLG refractive index is set to ñ=16.9+i49.2 as measured via THz-TDS (see Supplementary Material). **D**, Colormap of the waveguide loss spectrum for different values of the MLG refractive index imaginary part. The black dashed line is the maximum of the losses for each frequency. The MLG operate as a DGSA when the imaginary part of the refractive index is below this line. The purple dashed line is the imaginary part of the MLG refractive index as extracted from THz-TDS experiments. **E**, Simulated waveguide losses for TM00 and TM01 modes as a function of the imaginary part for Re(ñ$_g$)=16.9 at 3 THz. For Im(ñ$_g$)<65, with Im(ñ$_g$)=65 being the loss maximum at 3 THz, a reduction of Im(ñ$_g$), induced by saturable absorption, leads to a net round-trip gain which favours pulse formation.

Figure 2A shows the CW current density-voltage (J-V) and light-current density (L-J) characteristics of the DGSA-QCL. The threshold current density ($J_{th}$ ~170 A/cm$^2$) of the reference QCLs[49] proves that the DGSA architecture does not affect significantly $J_{th}$ (180 A/cm$^2$). The QCL delivers a maximum optical power (8 mW) and has a slope efficiency of 46 mW/A.

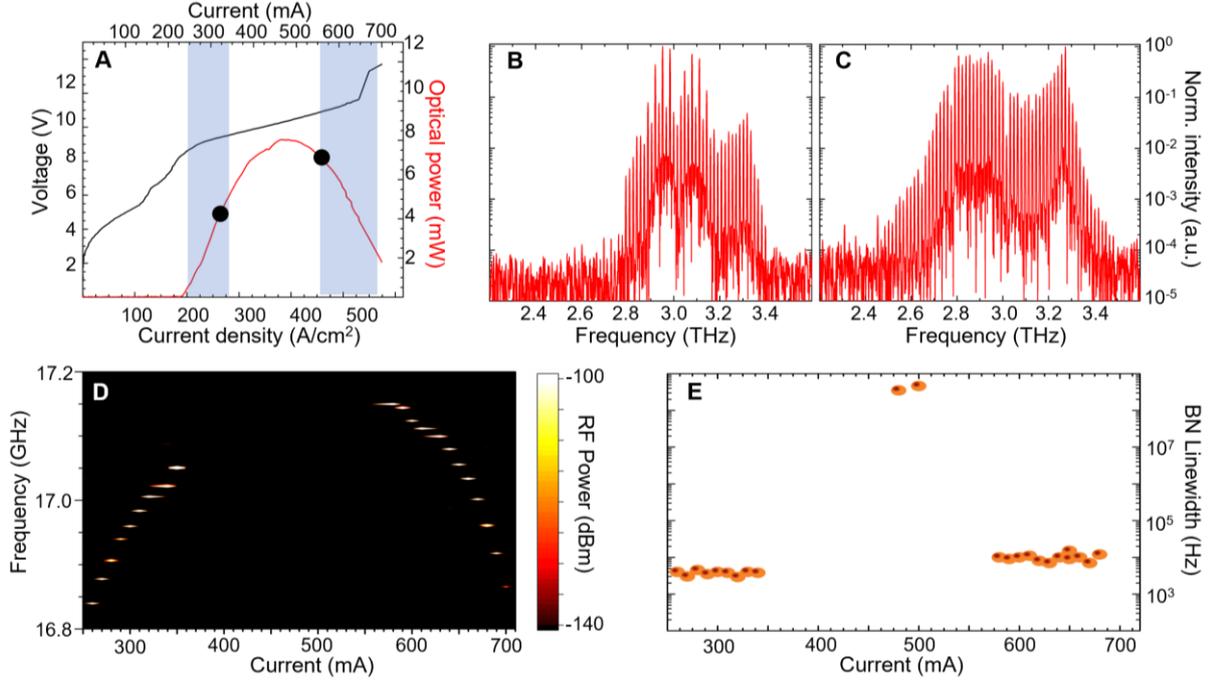

**Figure 2: Electrical and optical characterization of the DGSA-QCL**
**A,** Voltage−current density (V−J) and light−current density (L−J) characteristics measured at 20 K in continuous wave. The two shaded blue areas correspond to the bias ranges where the device shows a single and narrow intermode beatnote. The optical power was measured using a broad-area terahertz absolute power meter (TK Instruments, aperture 55 × 40 mm²). **B-C,** FTIR spectra acquired under vacuum at (**B**) 340 mA and (**C**) 570 mA at 20 K and with a 0.075 cm$^{-1}$ spectral resolution. **D,** Intermode beatnote map as a function of the DGSA-QCL driving current, measured at 20 K. **E,** Linewidths of the beatnotes in (**D**).

The Fourier transform infrared spectra (Bruker, Vertex 80) (Fig. 2B,C) collected under-vacuum above the whole band alignment (Fig. 2B) and in the region of maximum spectral coverage, above the peak optical power (Fig. 2C), show that the QCL bandwidth progressively increases from 0.9 to 1.25 THz. Furthermore, in these regimes, the modes appear phase-locked, as indicated by the sharp (~38 dBm) and narrow (1–5 kHz) intermode beatnotes (Fig. 2D,E), which provide a signature of a genuine comb operation. Generally, the emission of such a THz QCL operating as a FC is a mix of a frequency and amplitude modulated output[35], with a periodicity given by the cavity roundtrip.

The intermode beatnote map, plotted as a function of the driving current (Fig. 2D), shows that the comb regime in the DGSA laser persists at first for almost 100 mA, i.e. the bias region of the reference laser[49], and then immediately before and across the negative differential resistance (NDR) region. Conversely, the reference structure[49] behaves as a fully stabilized comb[59] only above the onset of band alignment and for a current range ~106 mA[49], regardless of cavity dimensions[49,60,61]. In fact, it is usually difficult to operate a QCL as a FC close to the NDR region since the high group velocity dispersion (GVD) occurring in the region of high electric field domains prevents the modes from achieving stable phase-locking[62,63]. In the current regime in-between (350–570 mA), the linewidth is >$10^8$ Hz, as in the reference structure for $J/J_{th}$ > 1.4[49,60], signature of a mostly chaotic behavior of the modal phases.

The analysis of the beatnote linewidths shows a general agreement with the values extracted on reference lasers (2–8 kHz)[49,60], whereas significantly smaller linewidths (600 Hz) are obtained when GVD compensation schemes are adopted[60,61]. This means that, as expected, the DGSA architecture does not affect the GVD and suggests that the appearance of a single and narrow beatnote at the NDR region is indicative of a different physical mechanism to that operated on the QCL by the GSA. Similar results (Supplementary Materials) have been achieved on a second DGSA-QCL with the same active region, 15 graphene layers and different cavity dimensions (3.13 mm × 68 μm × 17 μm), also with significant reduction of the beatnote linewidth in the NDR.

We next measured the emission profile of the DGSA in the time domain with a coherent technique, gaining access to the amplitude and phase information of the emitted electric field. To this aim, we performed a coherent measurement based on injection seeding[64], previously used to show pulsed behavior induced in THz QCLs by active mode-locking[64]. Such a technique allows measurement of the free-running QCL emission with a temporal resolution better than 100 fs.

Fig. 3A plots a 250-ps-long time-trace of the electric field emitted by the reference QCL[49] biased at 663 mA ($I/I_{th}$ = 1.70), acquired 1 ns after the seeding pulse injection, i.e. well into the steady-state regime where no effect of the initial pulse persists. The quasi-CW profile combined with amplitude modulations on a short time scale (<5 ps) is a typical consequence[65,66] of the interplay between the QCL fast gain recovery time[36] and the giant Kerr non-linearity[67] arising from Bloch gain[66]. The spectrum (Fig. 3B), obtained by Fourier-transforming the electric field time trace, shows broadband emission from 2.2 to 3.3 THz, in agreement with the FTIR spectra measured under-vacuum[49,60]. The latter, however, shows modes up to 3.45 THz[61] and an even power distribution between the two bands centered at 2.8 and 3.2 THz, meaning that the higher frequency band is more attenuated in the TDS spectrum than the band at lower frequencies. This is consistent with the presence of the absorption at 3.7 THz of the ZnTe crystal[68] used for coherent electro-optic sampling.

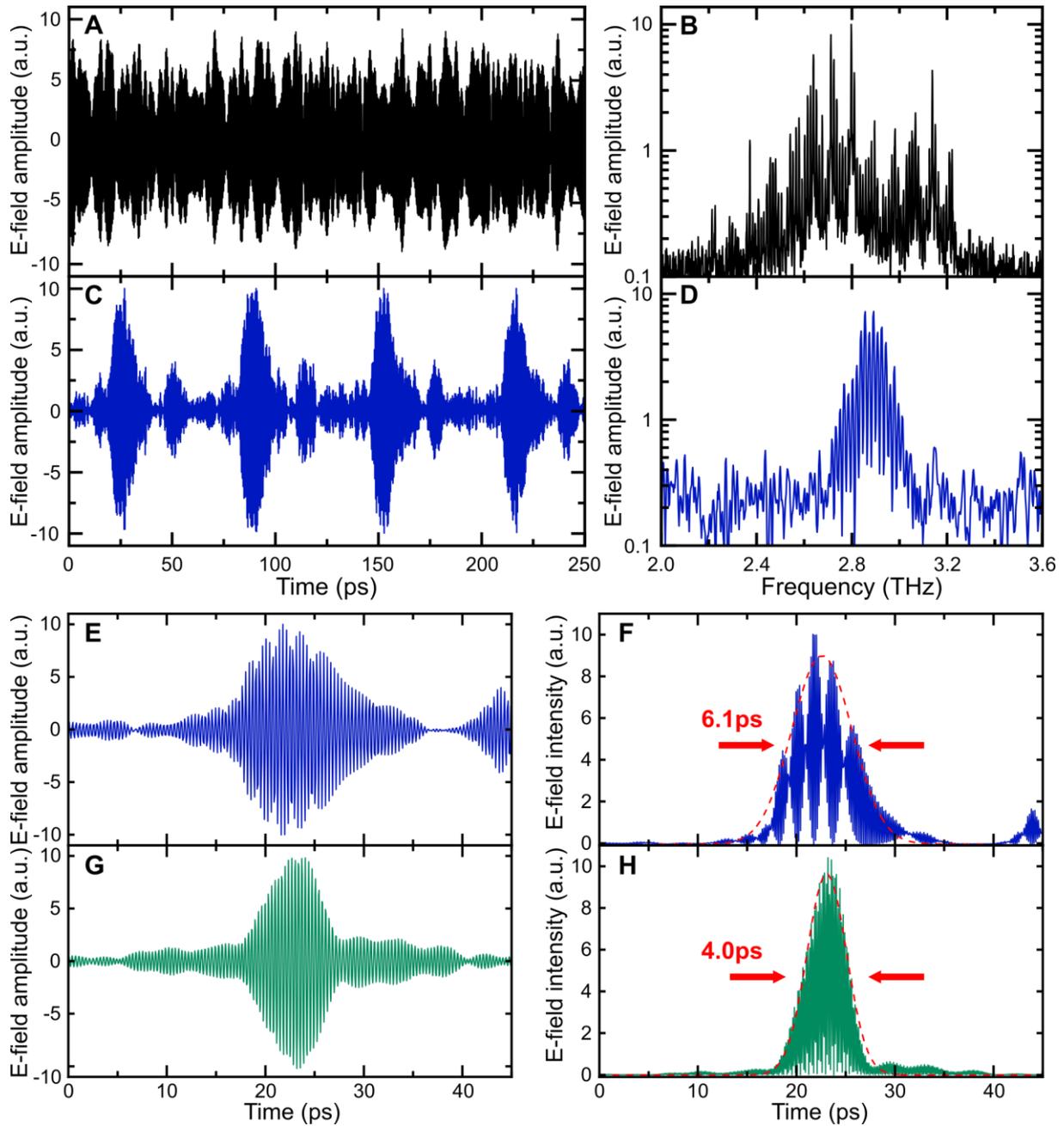

**Figure 3: THz-TDS emission profiles.**
**A,** Time-domain emission profile of the reference QCL (without GSA, Ref.[49]) biased at 663 mA. **B,** Fourier frequency transform (FFT) spectrum of the time trace in Fig. 3A. **C,** Time-domain emission profile of the DGSA-QCL biased at 327 mA. **D,** FFT spectrum of the time trace in Fig. 3C. **E,** Electric field amplitude profile of a pulse emitted by the DGSA-QCL at 327 mA. **F,** Intensity profile of the pulse in Fig. 3E. The red dashed line is a Gaussian fit of the pulse with full width at half maximum (FWHM) of 6.1 ps. **G,** Electric field amplitude profile of a pulse emitted by the DGSA-QCL at 570 mA. **H,** Intensity profile of the pulse in Fig. 3G. The red dashed line is a Gaussian fit of the pulse with FWHM of 4.0 ps.

Fig. 3C plots the free-running emission of the DGSA-QCL, biased at 327 mA (I/Ith =1.68) in the regime characterized by a 3 kHz-broad beatnote. The electric field profile reveals the generation of a pulsed profile, demonstrating that the QCL can be passively mode-locked by our GSA. The spectrum retrieved by fast-Fourier transforming the electric field (Fig. 3D) matches that measured by the FTIR at the same bias (Fig. 2B), aside for the attenuation at higher frequencies due to the ZnTe crystal absorption profile. Figure 3E is the electric field amplitude profile of one of the pulses emitted by the DGSA-QCL, biased at 327 mA, ~1 ns after the seeding pulse injection. The corresponding pulse intensity profile is in Fig. 3F, revealing a duration of ~ 6.1ps as extracted from a Gaussian fit (red dashed line). When the QCL is driven at 570 mA, where the number of emitted modes becomes significantly larger (Fig. 2C), the optical power per comb tooth reaches 0.12 mW and, with a sharp and narrow beatnote still present, and a pulsed behavior is once again retrieved (Fig. 3G). As expected, the broader bandwidth gets translated into a shorter pulse duration. This is evident from the intensity profile of the electric field in Fig. 3H, showing 4 ps pulse generation.

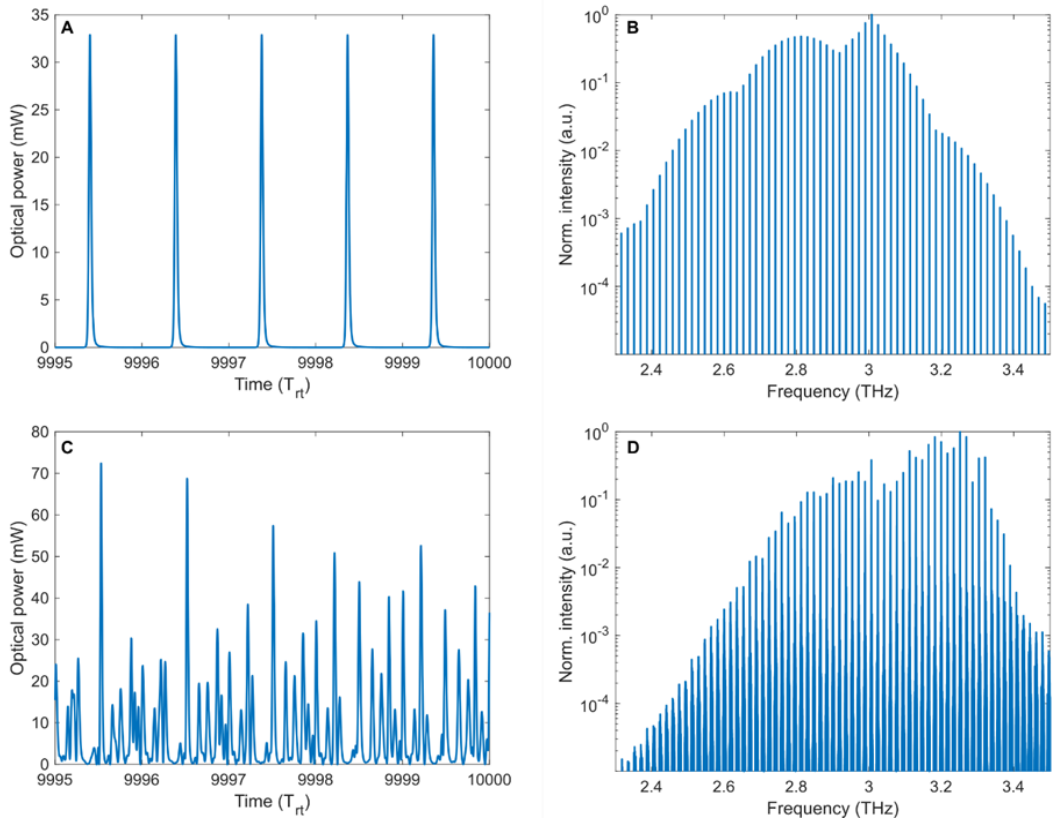

**Figure 4: Maxwell-Bloch dynamical simulations**.
**A,** Instantaneous pulse power outcoupled at the right end facet in steady state operation at ~ 2.7 times the threshold current for $P_s$=85mW. **B,** Fourier spectrum of the time trace in Fig. 4A. **C,** Outcoupled power with increased gain showing chaotic behavior. **D,** Fourier spectrum of the time trace in Fig. 4C.

To shed light on the physical mechanism at the origin of the observed phenomenon, we performed time-domain simulations[69] to study the pulse formation mechanism of our DGSA-QCL. The active region dynamics are modeled employing a Lindblad-type approach and the quantum system represents a QCL period with periodic boundary conditions[69,70] (Methods). The waveguide parameters used for the simulations are in Table 1. The simulated optical power of the last five roundtrip times ($T_{rt}$), out-coupled at the back facet in stationary operation and its Fourier transform are in Fig. 4A,B. A regular train of pulses with a duration (FWHM) ~1.7 ps is observed, consistent with the experimental results (4 ps).

The resulting comb shows clear, narrow (hundred Hz) and equidistant lines where 26 of them are in a 10dB power range. For comparison, enhancing the gain by increasing the doping density by ~4%, reproduces the experimental breakdown of pulses at currents 350 mA < I < 570 mA, i.e. in the current range above the breakdown of the initial comb state and before entering in the second stable comb regime, Fig. 4C,D, with the resulting chaotic behavior.

**Table 1. Simulation parameters.**
Waveguide parameters for our simulations.

| Parameter | Value |
|---|---|
| Loss coefficient $a_0$ | 10.5 cm$^{-1}$ |
| Saturable loss coefficient $a_1$ | 10.5 cm$^{-1}$ |
| Saturation power $P_s$ | 85 mW |
| Overlap factor | 0.97 |
| Center frequency | 3.2 THz |
| Mirror reflectivity (power) | 70.88 % |
| Refractive index | 3.6 |
| GVD factor $\beta_2$ | 1.0×10$^{-22}$ s$^2$m$^{-1}$ |

In order to prove quantitatively the presence of mode-locking in the simulated optical field, we employ power and phase noise quantifiers $M_{\sigma P}$ and $M_{\Delta\Phi}$,[71] as:

$$M_{\sigma P} = \frac{1}{N_{10}} \sum_{q=1}^{N_{10}} \sigma_{P_q} \qquad M_{\Delta\Phi} = \frac{1}{N_{10}} \sum_{q=1}^{N_{10}} \sigma_{\Delta\Phi_q}$$

with

$$\sigma_{P_q} = \sqrt{\langle (P_q(t) - \langle P_q(t)\rangle)^2 \rangle} \qquad \sigma_{\Delta\Phi_q} = \sqrt{\langle (\Delta\Phi_q(t) - \langle \Delta\Phi_q(t)\rangle)^2 \rangle}$$

where $P_q(t)$ are the modal amplitudes and $\Delta\Phi_q(t)$ the modal phase differences between one mode and the adjacent, at time $t$ for each mode q = 1, …, $N_{10}$, with $N_{10}$ being the number of modes in the -10dB spectral bandwidth.

In this framework, an optical FC is present for $M_{\sigma P} < 10^{-2}$ mW and $M_{\Delta\Phi} < 2\times10^{-2}$ rad. Regarding the pulsed mode of Fig. 4A,B, values of $2\times10^{-8}$ mW and $2\times10^{-6}$ rad are obtained. We can thus confirm an ultra-stable mode-locked operation in this current range. Calculating the same parameters for the results of Fig. 4C,D, yields $M_{\sigma P} = 0.21$ mW and $M_{\Delta\Phi} = 9.34$ rad. This increase proves the absence of mode-locking for higher gain, in agreement with our theory of pulse formation, as discussed below.

**Discussion**

One key observation to understand pulse formation in our DGSA-QCL is that the graphene stripes provide quasi-instantaneous saturable absorption across the resonator length, while also the gain has a recovery time much faster (a few ps) than the roundtrip time[34]. Ref.[72] suggests that mode-locked operation only occurs if the pulses are stabilized by a net gain window, while the standard passive mode locking theory[72] does not provide self-starting mode-locked pulse solutions for instantaneous gain and saturable absorption. These assumptions are relaxed in our experiments by the fact that the gain, despite its fast recovery, is still slower than the saturable absorption provided by graphene[73,74]. Thus, a narrow gain region can exist slightly above the threshold gain where self-starting mode-locked pulse operation occurs. As also pointed out in[72], the constraints for self-starting passive mode-locking are further relaxed if a second, longer gain recovery time is present, as is typically the case for bound-to-continuum QCLs due to the electron transport across the miniband[75]. Deviations from the analytical theory of Ref.[72] arise for QCLs. In particular, the gain saturation and GSA bleaching due to the propagating pulse open up a short net gain window stabilizing the pulse, but introduce a net loss for counterpropagating fields, which helps suppress the buildup of a CW background. Furthermore, quantum coherence and nonlinear effects, spatial hole burning, as well as considerable outcoupling at the end facets, have to be taken into account necessitating our detailed Maxwell-Bloch-type simulations presented in Methods.

Aside from the experimental non-idealities, the main difference between simulated and experimental pulses can be ascribed to the available number of spectral modes and their power distribution, which is considerably more uniform in the simulations. This suggests that the main limitation in terms of achievable pulse duration is the laser bandwidth. Further optimization of the waveguide structure, adjustments to the graphene optical properties and the employment of ultra-broadband (an octave) active regions could be adopted to enhance the emission bandwidth uniformity, effectively indicating a practical path to achieve sub-ps THz pulse generation.

**Conclusions**

We have shown ultrashort pulse generation from a passively modelocked semiconductor THz laser using a unique graphene based distributed saturable absorber. Exploiting the wide spectral gain of THz QCLs and the lithographic capability to embed miniaturized, highly nonlinear, intra-cavity graphene absorbers to modulate light and produce short pulses without any external source or seeding system, could allow QCLs to become an attractive alternative to present-day THz TDS systems in spectroscopic and imaging applications. This will advance quantum science enabling users to capture the ultrafast dynamics in novel material systems, even down to the nanoscale[76], or performing real-time pulsed imaging and time of flight tomography[77], or time-resolved spectroscopy of gases, complex molecules and cold samples[78]. Other opportunities include the coherent control of quantum systems[79], quantum sensing and metrology where laser excitation can match the energy level splitting of molecules (and its pulsed nature can down-convert the spectrum to the RF domain[80]), or for ultra high-speed THz communications[44,81], greatly expanding the technological impact of QCLs.

**Methods**

**Fabrication and graphene preparation**

Fabry-Perot laser bars are fabricated in a metal–metal waveguide configuration via Au-Au thermo-compression wafer bonding of a17-μm-thick active region onto a highly doped GaAs substrate. This is followed by the removal, through a combination of mechanical lapping and wet etching, of the host GaAs substrate of a molecular-beam-epitaxially-grown (MBE) material. An $Al_{0.5}Ga_{0.5}As$ etch stop layer is then removed using HF etching. Vertical sidewalls are defined by inductively coupled plasma etching of the laser bars to provide uniform current injection. A Cr/Au (10 nm/150 nm) top contact is then deposited along the center of the ridge surface, leaving a thin region uncovered along the ridge edges and two stripes uncovered 12 μm from the center of the cavity. 3-μm-wide Ni (5-nm-thick) side absorbers were then deposited over the lateral uncovered region using a combination of optical lithography and thermal evaporation. These lossy side absorbers are intended to inhibit lasing of the higher order lateral modes by increasing their threshold gain[82].

The multilayer graphene sample is prepared and transferred on the stripes, using a wet transfer technique: A4-950K ply(methyl-methacrylate) polymer (PMMA) is spin coated at 2000 rpm on the surface of a single layer (SLG) sample (1cm ×1cm) grown on Cu via CVD. After 1 minute on a hot plate at 90°C, the sample is placed in a solution of 1 g of ammonium persulfate and 40 ml of DI water to etch the Cu substrate. Once the Cu etching is complete, the PMMA-SLG film is transferred in a beaker with DI water and then lifted with a second Cu-graphene square to obtain a bilayer graphene sample. This sample is left to dry overnight and finally the PMMA is removed with acetone. The Cu of the bilayer graphene is etched with the same technique, and then lifted by another SLG on Cu. This process is repeated until the desired MLG thickness is reached, in our case seven layers. The MLG is transferred onto the QCL top contact and then removed from the sides of the laser cavity and from the top Au contact by oxygen reactive ion etching (RIE).

The backside of the substrate is then lapped down to 150 μm for thermal management and enable operation in CW. Laser bars, 50 μm wide and 2.2 mm long, are then cleaved and the device mounted on a copper bar, wire bonded, and then mounted onto the cold finger of a He continuous-flow cryostat.

**Model**

The active region dynamics is modeled based on a Lindblad-type approach, where the quantum system represents a QCL period with periodic boundary conditions[69,70]. We performed self-consistent carrier transport simulations of the active region to extract the corresponding eigenenergies, optical dipole moments as well as the scattering and dephasing rates[69]. For simplicity, the dynamical model only takes into account the 2.5 THz design, yielding agreement with experiments. The model system comprises 9 quantized quantized energy states, including two upper- and one lower-laser levels. In the common rotating wave/slowly varying amplitude approximation, the optical cavity field is described by complex amplitudes $E^{\pm}(z,t)$ for the forward and backward traveling electric field component, and the propagation given by[70]:

$$v_g^{-1}\partial_t E^{\pm} \pm \partial_z E^{\pm} = p^{\pm} - a(P)E^{\pm} - i\frac{\beta_2}{2}\partial_t^2 E^{\pm} \qquad (1)$$

Here, $z$ and $t$ are the propagation coordinate and time, $v_g$ denotes the group velocity, $p^{\pm}(z,t)$ is computed from the Lindblad equation[69,70] and contains the polarization due to the quantized states of the QCL active region, and $\beta_2$ describes the background group velocity dispersion. The quasi-instantaneous SA is modeled by the power loss coefficient[83]:

$$a(P) = a_0 + \frac{a_1}{1+P/P_s} \qquad (2)$$

with the optical power $P \propto |E^+|^2 + |E^-|^2$. Electromagnetic simulations yield $a_0 = a_1 = 10.5$ cm$^{-1}$ (Fig. 1E). The saturation power $P_s$ depends on the electric field strength and orientation in the graphene stripes, hence it does not correspond to the saturation intensity of the MLG itself as it could be retrieved through z-scan measurements or from considerations on the number of layers and doping level, but rather to the effective value for the transverse waveguide mode. Thus, we use $P_s$ as the only fitting parameter in our simulation. The resulting Maxwell-Bloch-type model is numerically solved over ten thousand roundtrips to ensure convergence to steady state[69].

**Acknowledgements** This work was supported by the European Research Council through the ERC Consolidator Grant (681379) SPRINT (MSV), FET Open project EXTREME IR (944735) (MSV, SSD), ERC GSYNCOR (ACF), HETERO2D (ACF) , EIC CHARM (ACF), the French National Research Agency (ANR-18-CE24-0013-02 - "TERASEL") (SD), and the EPSRC (UK) programme grant 'HyperTerahertz' (EP/P021859/1) (EHL, LL, AGD), EP/L016087/1, EP/K01711X/1, EP/K017144/1, EP/N010345/1, EP/V000055/1 (ACF), and Graphene Flagship (MSV, ACF)

**Author contributions:**

M.S.V. and V.P. conceived the concept. E.R. fabricated the devices, set up the transport and optical experiment. V.P. performed numerical simulations and interpreted the data. E.R. S.S.D., S.K. A.D.V.

acquired the experimental data. L.L, A.G.D. and E.H.L grew by molecular beam epitaxy the QCL structure. C.J. and L. L. developed the theoretical model. The manuscript was written by M.S.V. and V.P. M.S.V. coordinated and supervised the project. All authors contributed to the final version of the manuscript and to discuss the results.

**Supporting Information Available**

The following files are available free of charge. SupportingInfo.pdf

**Competing financial interests:**

The authors declare no competing financial interests.

**Data and materials availability** The data presented in this study are available on reasonable request from the corresponding author.